# Study of Group III-V Waveguides on Sapphire Platform for Photonic Integrated Circuits

**Manoj Kumar Shah [1], Richard A. Soref [2], Diandian Zhang [3], Wei Du [3,4,*], Gregory J. Salamo [3,4], Shui-Qing Yu [3,4] and Mansour Mortazavi [1]**

[1.] Department of Chemistry and Physics, University of Arkansas at Pine Bluff, Pine Bluff, AR 71601, USA;

[2.] Department of Engineering, University of Massachusetts at Boston, Boston, MA 02125, USA

[3.] Department of Electrical Engineering and Computer Science, University of Arkansas, Fayetteville, AR 72701, USA

[4.] Institute for Nanoscience and Engineering, University of Arkansas, Fayetteville, AR 72701, USA

**\*** Correspondence: weidu@uark.edu

**Abstract:** Photonic integrated circuits (PICs) have been acknowledged as the promising platforms for the applications in data communication, Lidar in autonomous driving vehicles, innovative sensor technology, etc. Since the demonstration of optical components individually, integration of both electronics and photonics for functional devices on a common platform has been a key technology driver enhancing the stability and scalability of integrated photonic technologies. Recently, we proposed to use sapphire as a high-performance PIC platform, which enables a fully integrated solution to include a complete set of components with light source, modulator, light detection, passive devices, silicon on sapphire control circuit all-in-one sapphire platform to achieve high-performance low-cost mixed-signal optical links. In parallel to developing active components such as group III-V lasers on sapphire, in this work, the performance of group III-V straight waveguides on sapphire was systemically studied. The refractive indices contrast between GaAs, InP, GaSb, and sapphire are sufficiently high to achieve low loss over a broad optical wavelength. The calculated loss at wavelengths of 1330 nm, 1550 nm, and 2000 nm for the GaAs, InP, and GaSb rib waveguides are 0.32 dB/cm, 0.67 dB/cm, and 0.70 dB/cm, respectively. Since the fundamental element to construct all passive building blocks is the straight waveguide, results from this work would allow us to assess other basic passive building blocks.

## 1. Introduction

The integration of two or more optical devices to generate, manipulate, transport, and measure optical signals, is widely known as photonic integrated circuits (PICs) [1-3]. PIC has major advantages over conventional integrated circuits (ICs) that include high data rate, reduced propagation loss, and most importantly, orders of magnitude wider bandwidth [4-7]. Photonics provides a powerful capability of achieving complicated signal processing that the conventional IC could not easily obtain [8, 9]. Nowadays PIC has penetrated to a variety of civilian applications such as cellular, wireless, satellite communications, cable television, distributed antenna systems, optical signal processing and medical imaging systems. The future driving forces for PIC are expected to be broadband wireless access networks, the demand for more efficient wireless infrastructures due to the blooming of mobile device platforms, future wearable electronics, and emerging direct fiber link to home and in-home networks [10-16].

Currently, PIC is mainly implemented in the following material platforms: i) the InP platform inherently supports light generation, amplification, modulation, detection, and switching in addition to passive functionalities [1, 17]. InP photonics is highly attractive for large scale photonic integration and the PICs are highly complex. However, it is well known that the propagation losses in InP optical waveguides can be an order of magnitude higher compared to waveguides based on silica or silicon. For some applications, this large propagation loss needs to be compensated by optical gain from active components like semiconductor optical amplifiers which could result in adding noise and limiting the dynamic range. ii) Silicon photonics is one of the most exciting and fastest growing photonic technologies in recent years [2]. The initial pull of this technology is its compatibility with the mature silicon IC manufacturing. Another motivation is the availability of high-quality SOI planar waveguide circuits that offer strong optical confinement due to the high index contrast between silicon and $SiO_2$. However, due to the indirect bandgap of Si, Si photonics utilizes III-V materials for the light source. iii) TriPleX™ technology is based on a combination of $Si_3N_4$ as waveguide layer, filled by and encapsulated with $SiO_2$ as cladding layers [3]. This technology allows for extremely low loss integrated optical waveguides both on silicon and glass substrates for all wavelengths in between 405 nm up to 2.35 µm, providing maximum flexibility from an integration standpoint. Since no light sources, detectors, amplifiers, and modulators are available, this technology therefore requires a hybrid approach with InP.

Ideally, a PIC with simultaneous photonics, electronics, and complementary metal-oxide–semiconductor (CMOS) compatibility is the target. Here we propose a sapphire-based PIC platform which enables the monolithic integration of optically advantageous group III-V semiconductors on a single sapphire substrate. Specifically, we propose a III-V semiconductor laser, waveguide, and optical amplifier, epitaxially grown on a sapphire substrate. The rationale is based on the following factors: i) sapphire is more favorable for growth of III-V materials compared with Si substrates due to its closely matched coefficient of thermal expansion (CTE) with III-V materials including GaAs, InP, and GaSb, etc (Table I). As a result, the high-performance III-V lasers on the sapphire platform can be expected. i) sufficiently low refractive index of sapphire enables the low loss waveguides and other passive devices. ii) the sapphire platform is not limited by strong two-photon absorption at 1.55 µm, currently a serious problem for Si photonics [18]. iii) Silicon-on-sapphire wafers are available from foundry service, which allows for the seamless integration of photonics components with standard CMOS circuitry.

Table 1. Material Properties of III-V Photonic Integrated Circuits.

| Materials | Thermal conductivity (W/m/K) | Coefficient of thermal expansion | Wavelength of interest (µm) | Refractive index | Bandgap (eV) |
|---|---|---|---|---|---|
| GaAs | 46 | 5.73 | 0.8 – 2.4 | 3.6035 – 3.3479 | 1.39 |
| AlAs | 41 | 5.2 | 0.8 – 2.4 | 3.0144 – 2.8652 | 2.16 |
| InP | 68 | 4.5 | 0.9 – 2.5 | 3.4415 – 3.116 | 1.344 |
| GaP | 110 | 6.3 | 0.9 – 2.5 | 3.2092 – 3.1043 | 2.26 |
| GaSb | 34 | 7.4 | 1.9 – 3.5 | 3.7903 – 3.7248 | 0.67 |
| AlSb | 60 | 4.10 | 1.9 – 3.5 | 3.2192 – 3.1672 | 1.16 |
| Sapphire | 35 | 5.5[c] | 0.8 – 3.5 | 1.7601 – 1.6953 | 9.0 |
| Si | 148 | 2.6 | 0.8 – 3.5 | 3.6941 – 3.4265 | 1.12 |

## 2. Simulation Methods

The rib and strip waveguide structures were simulated to investigate the performance of III-V waveguides on a sapphire substrate, as shown in Figure 1 (cross-sectional view). The GaAs, InP, and GaSb waveguide core layers were studied. For each structure, a wetting layer was inserted between the core layer and the sapphire substrate to reproduce the practical material growths, which are AlAs, GaP, and AlSb for GaAs, InP, and GaSb waveguides, respectively. The wetting layers aim to increase the core layer coverage on the sapphire substrate and to effectively reduce the twining defects. Experimental results confirmed that the presence of AlAs improves the wetting of the substrate compared to the direct growth of GaAs [19]. The entire device is surrounded by air.

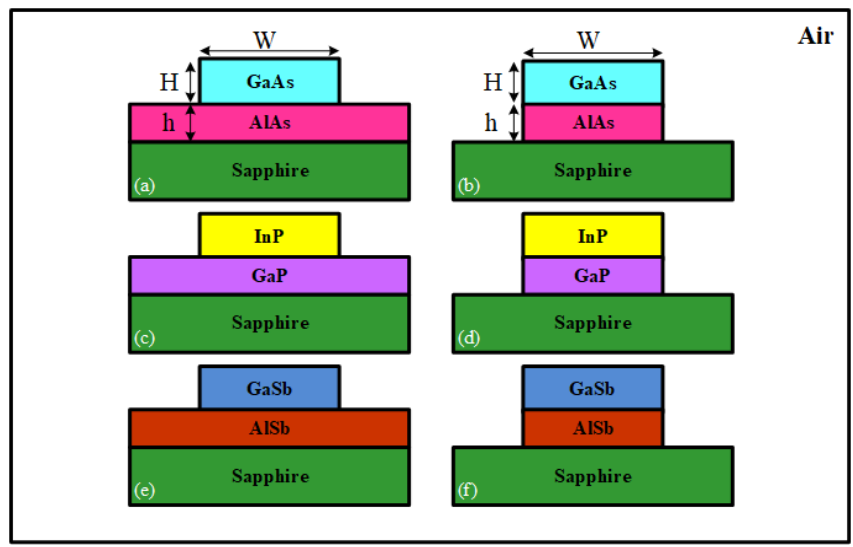

**Figure 1.** Cross-sectional view of III-V-on-sapphire rib and strip waveguides for (a) and (b) GaAs/AlAs, (c) and (d) InP/GaP, (e) and (f) GaSb/AlSb.

For the simulation of the guided modes in the waveguide, three aspect ratios of the width (W) and height (H) of the core layer were considered, i.e., W = 1.5H, 2.0H, and 2.5H. The thickness of the wetting layer (h) was kept at 40 nm for all cases. For the rib waveguides, the widths of wet layers were set as 10 µm. While for the strip waveguides, the widths of wet layers were the same as core layer widths. The complex refractive indices of core materials GaAs, InP, and GaSb, and corresponding wetting layer materials were taken from references [20-22], including the real part of the refractive index (n) and the attenuation index (k). The wavelength of interest for GaAs, InP and GaSb waveguides covers the spectra of corresponding light emitters, from visible to the shortwave infrared regime. Commercial software Ansys Lumerical was used to extract the effective index and power confinement factor of the guided mode in waveguides by effective index method using Ansys Lumerical Mode's tool, followed by the parametric sweep being performed to study the impact of varying dimension of waveguide on single-mode, multi-mode and cut-off conditions. The propagation loss of single mode waveguides was calculated by estimating the attenuation coefficient incurred due to material absorption loss, optical scattering loss at wet layer/substrate interface and sidewall roughness, and radiation leakage.

## 3. Results and discussions

Figures 2(a-f) show a guided mode colormap of the simulated fundamental TE and TM polarizations for the III-V rib waveguide structures. The colormap represents the power confinement factor (PCF) in the core of the waveguide. The color bars of each figure correspond to the colormap data. A broad wavelength range in NIR and MIR spectral windows is guided by varying the core width from 200 nm to 1100 nm and maintaining the aspect ratio of W = 1.5H. The regions of multi-mode, single-mode, and cut-off are given in Figure 2(a, c, and e) for TE polarizations, and in Figure 2(b, d, and f) for TM polarizations of rib waveguides. The single-mode propagation with a good PCF can be guided in the core of the waveguide for all three materials. The insets in Figure 2(a-f) show the waveguide mode intensity profile corresponding to the white star marked in the guided mode colormaps. The waveguide width for single-mode propagation of GaSb is the smallest followed by InP then GaAs, contrary to the refractive index differences of core-sapphire. This is predictable because the larger refractive index difference of core-sapphire leads to stronger confinement, as shown in Figure 2. Besides, from Figure 2 we see that the TE polarization [Figure 2(a, c, and e)] is guided over a wider wavelength range than the TM polarization [Figure 2(b, d, and f)]. This is due to the lower refractive index differences at in-plane than out-of-plane [23].

The power confinement factor in a waveguide depends on the refractive index contrast between the core-cladding regions, operating wavelength, as well as the geometric dimension of the waveguides [24]. From Figure 2 we see for the same geometric dimension and wavelength,

GaSb waveguides have the highest PCF, followed by InP waveguides and then GaAs waveguides. This is due to the highest refractive index contrast at GaSb-sapphire, followed by InP-sapphire and then GaAs-sapphire regions.

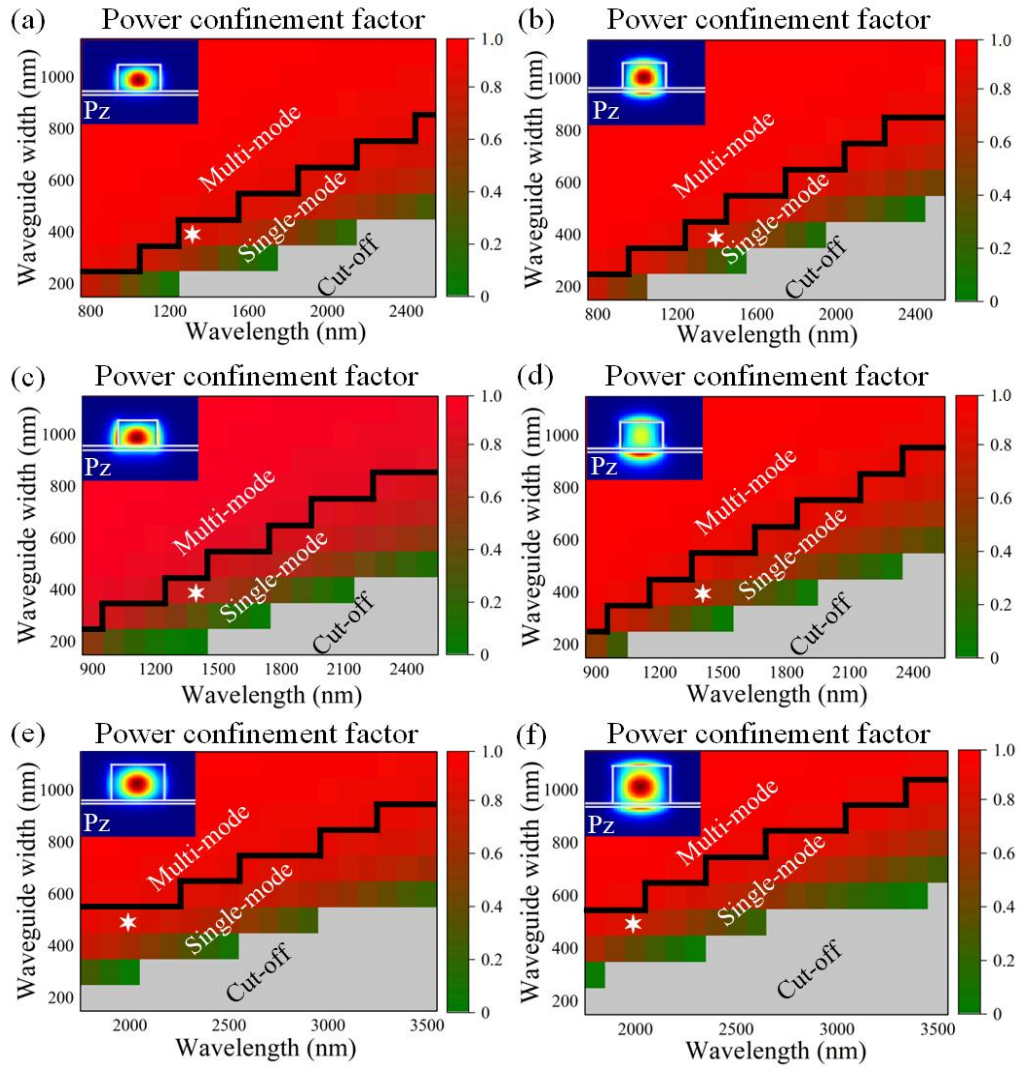

**Figure 2.** Simulated guided-mode map for rib waveguides with the general structure shown in Fig. 1 when W = 1.5H, h = 40 nm and for the following cases: (a) TE polarization, GaAs/AlAs waveguide, (b) TM polarization, GaAs/AlAs waveguide, (c) TE polarization, InP/GaP waveguide, (d) TM polarization, InP/GaP waveguide, (e) TE polarization, GaSb/AlSb waveguide, and (f) TM polarization, GaSb/AlSb waveguide. In each figure, the colormap represents the power confinement factor in the core of waveguides. The regions of multi-mode, single-mode, and cut-off have been highlighted. The insets in a-f show the waveguide mode intensity profile corresponding to the white star marked in the guided-mode map.

The GaAs, InP, and GaSb strip waveguides were analyzed using the same simulation approach, as shown in Figs. 3(a-f). For TE and TM polarizations the aspect ratio of W = 1.5H was maintained. The guided modes are similar to that of rib waveguides as seen in Figure 2, however the PCF in the rib waveguide is slightly higher than in strip waveguides at the same wavelength and dimensions. This is due to the rib structure providing additional optical confinement of the light in the vertical direction [25]. In the strip waveguides, the mode size is smaller than in rib waveguides as a result of the stronger light confinement in the horizontal direction, which leads to the optical field penetrating into the substrate and consequently the higher propagation loss of the strip waveguides.

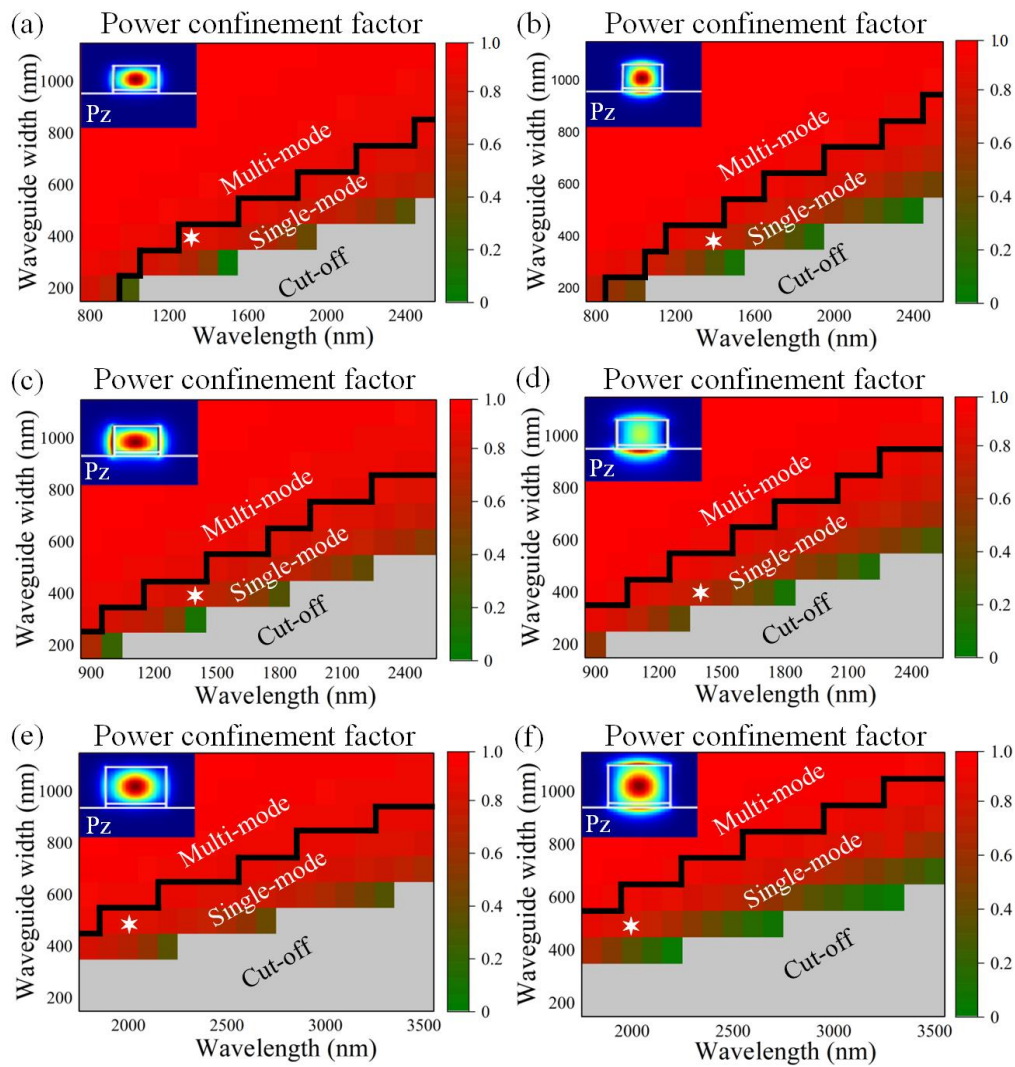

**Figure 3.** Simulated guided-mode map for strip waveguides with the general structure shown in Fig. 1 when W = 1.5H, h = 40 nm and for the following cases: (a) TE polarization, GaAs/AlAs waveguide, (b) TM polarization, GaAs/AlAs waveguide, (c) TE polarization,

InP/GaP waveguide, (d) TM polarization, InP/GaP waveguide, (e) TE polarization, GaSb/AlSb waveguide, and (f) TM polarization, GaSb/AlSb waveguide. In each figure, the colormap represents the power confinement factor in the core of waveguides. The regions of multi-mode, single-mode, and cut-off have been highlighted. The insets in a-f show the waveguide mode intensity profile corresponding to the white star marked in the guided-mode map.

To analyze the single-mode condition, a straight-line approximation was applied for rib and strip waveguides as $W = m\lambda + c$ (where W is waveguide width, $\lambda$ is a wavelength, and m and c are constants), which allows direct comparison of single-mode width with the operating wavelength. Figure 4 plots the single-mode conditions of III-V-on-sapphire substrate for rib and strip waveguides in the NIR and MIR regimes. The single-mode condition in the plot corresponds to a power confinement factor close to or above 80%. It can be seen that the single-mode width of rib waveguides is wider than that of strip waveguides for all the cases (W = 1.5H, 2.0H, and 2.5H). This is due to the mode size and higher light confinement in the core of rib waveguides than in strip waveguides. Single-mode operation of GaSb waveguides covers the broadest wavelength range compared to InP and GaAs waveguides.

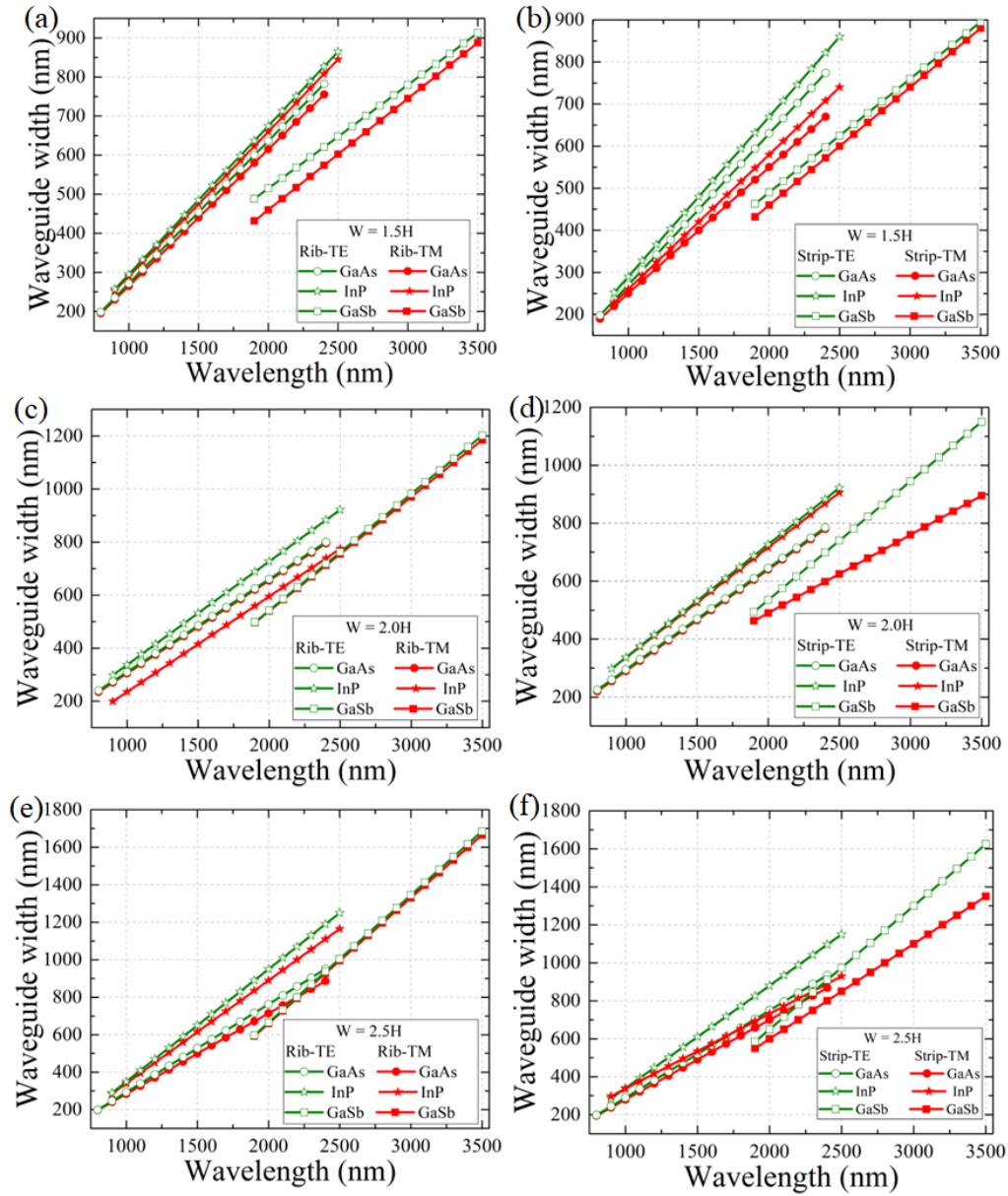

**Figure 4.** A straight-line approximation showing the single-mode condition for TE and TM polarization modes for rib and strip waveguides for the case: (a) W = 1.5H, rib, (b) W = 1.5H, strip, (c) W = 2.0H, rib, (d) W = 2.0H, strip, (e) W = 2.5H, rib, and (f) W = 2.5H, strip. The single-mode condition in the plot corresponds to the power confinement factor close to or above 80%.

**4. Loss analysis of III-V waveguides on sapphire**

The propagation loss ranging between 0.19 dB/cm and 4 dB/cm in III-V semiconductor straight waveguides has been reported [26-29]. The loss mechanisms in the semiconductor waveguides are mainly material absorption loss due to free carrier absorption, optical scattering loss due to some dislocation at the nanoscale wet layer/substrate interface and sidewall roughness, and radiation leakage [28, 30].

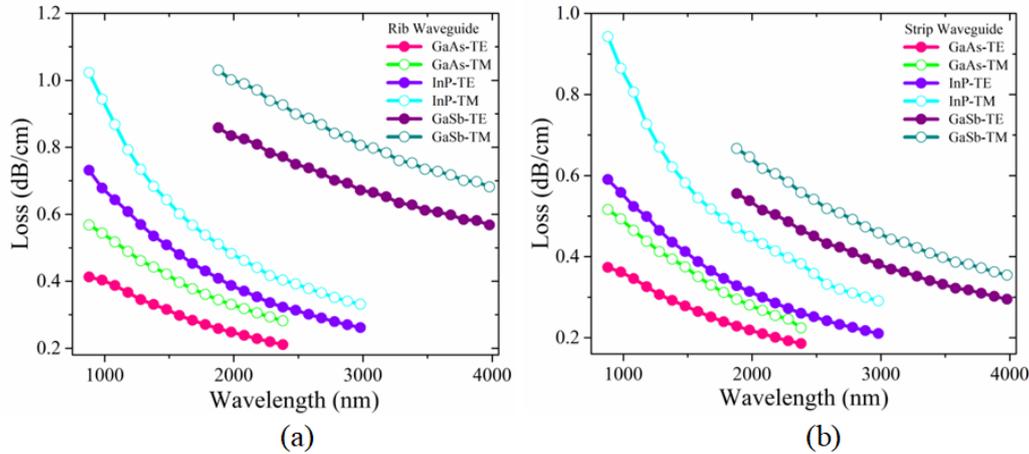

**Figure 5.** Simulated propagation losses of the GaAs/AlAs-on-sapphire single-mode rib and strip waveguides for TE and TM polarization modes over the broad wavelength spectrum from 800 nm to 3500 nm.

The propagation loss of fundamental TE and TM polarizations for the rib and strip waveguides structures were calculated for GaAs, InP, and GaSb materials. The losses are estimated for the waveguides with an aspect ratio of W = 1.5H which features larger mode size and higher optical confinement. The core width is varied between 300 nm and 900 nm for GaAs and InP waveguides, and for GaSb, width is varied between 400 nm and 1000 nm (estimated from the single mode approximation). The free carrier < $10^{15}$ cm$^{-3}$ in the core for bulk materials yields absorption loss ≤ 0.01 dB/cm [24], which is negligible in the simulations. The propagation losses for rib waveguides are significantly smaller than for strip waveguides [31]. The scattering loss due to sidewall roughness for rib and strip waveguides was estimated as to be 0.05 dB/cm and 0.25 dB/cm, respectively, and loss due to dislocations at nanoscale wet layer/substrate interface was 10 dB/cm [31, 32]. The attenuation index (k) for the losses due to sidewall roughness and dislocations at the buffer/substrate interface are calculated by using k = 4.3 (λα/4π), where λ and α are the operating wavelength and attenuation constant in dB/cm, respectively. The propagation losses for TE and TM polarization for rib and strip waveguides over the broad spectral range between 800 nm and 3500 nm is depicted in Fig. 5. The propagation losses for the rib waveguide, as expected, are smaller than for the strip waveguide. The losses for the TE polarization are smaller than for TM polarization over the entire operating wavelength, this is due to lower refractive index difference at in-plane than at out-of-plane. The estimated loss for commonly used wavelengths 1330 nm, 1550 nm, and 2000 nm for the GaAs, InP, and GaSb rib waveguides are 0.32 dB/cm, 0.67 dB/cm, and 0.70 dB/cm, respectively. These estimated losses are comparable to the reported III-V waveguide losses on a silicon wafer [26-28].

## 5. Conclusions

The rib and strip waveguides of three materials (GaAs, InP, GaSb) on sapphire are analyzed. Tightly built rib and strip waveguides with low losses were proved. The estimated loss for commonly used wavelengths 1330 nm, 1550 nm, and 2000 nm for the GaAs, InP, and GaSb rib waveguides are 0.32 dB/cm, 0.67 dB/cm, and 0.70 dB/cm, respectively. These estimated losses are comparable to the reported III-V waveguide losses on silicon wafer. These properties demonstrate that the III-V waveguide on sapphire have good performance, which is acceptable for PICs. Considering that the high-performance III-V active devices such as lasers and detectors can be built on the same sapphire substrate, the PICs on sapphire could stand out from the rest of material systems.

**Author Contributions:** Simulation, writing and editing, Manoj Kumar Shah; Waveguide design, and supervising, Richard A. Soref; Simulation and editing, Diandian Zhang; Simulation direction and paper revision, Wei Du; Paper revision, Gregory J. Salamo; Supervising and revision, Shui-Qing Yu; Supervising and funding, Mansour Mortazavi.

**Funding:** This work is supported by the Air Force Office of Scientific Research and Air Force Research Laboratory under agreement W911NF-20-1-0270, the Department of Energy – National Nuclear Security Administration under agreement DE-NA0004114, the National Science Foundation under grant number 2327229, and the Chancellor's Innovation Commercialization Fund from University of Arkansas (WFCSF Commercialization FY2023-15).